\documentclass[iop]{emulateapj}
\usepackage[]{hyperref}
\usepackage{textcomp}
\usepackage{graphicx}
\usepackage{amsmath}
\usepackage{amssymb,latexsym}
\usepackage{threeparttable}
\def\jnl@style{\it}
\def\aaref@jnl#1{{\jnl@style#1}}

\def\aaref@jnl#1{{\jnl@style#1}}

\def\aj{\aaref@jnl{AJ}}                   
\def\araa{\aaref@jnl{ARA\&A}}             
\def\apj{\aaref@jnl{ApJ}}                 
\def\apjl{\aaref@jnl{ApJ}}                
\def\apjs{\aaref@jnl{ApJS}}               
\def\ao{\aaref@jnl{Appl.~Opt.}}           
\def\apss{\aaref@jnl{Ap\&SS}}             
\def\aap{\aaref@jnl{A\&A}}                
\def\aapr{\aaref@jnl{A\&A~Rev.}}          
\def\aaps{\aaref@jnl{A\&AS}}              
\def\azh{\aaref@jnl{AZh}}                 
\def\baas{\aaref@jnl{BAAS}}               
\def\jrasc{\aaref@jnl{JRASC}}             
\def\memras{\aaref@jnl{MmRAS}}            
\def\mnras{\aaref@jnl{MNRAS}}             
\def\pra{\aaref@jnl{Phys.~Rev.~A}}        
\def\prb{\aaref@jnl{Phys.~Rev.~B}}        
\def\prc{\aaref@jnl{Phys.~Rev.~C}}        
\def\prd{\aaref@jnl{Phys.~Rev.~D}}        
\def\pre{\aaref@jnl{Phys.~Rev.~E}}        
\def\prl{\aaref@jnl{Phys.~Rev.~Lett.}}    
\def\pasp{\aaref@jnl{PASP}}               
\def\pasj{\aaref@jnl{PASJ}}               
\def\jsara{\aaref@jnl{JSARA}}             
\def\qjras{\aaref@jnl{QJRAS}}             
\def\skytel{\aaref@jnl{S\&T}}             
\def\solphys{\aaref@jnl{Sol.~Phys.}}      
\def\sovast{\aaref@jnl{Soviet~Ast.}}      
\def\ssr{\aaref@jnl{Space~Sci.~Rev.}}     
\def\zap{\aaref@jnl{ZAp}}                 
\def\nat{\aaref@jnl{Nature}}              
\def\iaucirc{\aaref@jnl{IAU~Circ.}}       
\def\aplett{\aaref@jnl{Astrophys.~Lett.}} 
\def\apspr{\aaref@jnl{Astrophys.~Space~Phys.~Res.}}
\def\bain{\aaref@jnl{Bull.~Astron.~Inst.~Netherlands}} 
\def\fcp{\aaref@jnl{Fund.~Cosmic~Phys.}}  
\def\gca{\aaref@jnl{Geochim.~Cosmochim.~Acta}}   
\def\grl{\aaref@jnl{Geophys.~Res.~Lett.}} 
\def\jcp{\aaref@jnl{J.~Chem.~Phys.}}      
\def\jgr{\aaref@jnl{J.~Geophys.~Res.}}    
\def\jqsrt{\aaref@jnl{J.~Quant.~Spec.~Radiat.~Transf.}}
\def\memsai{\aaref@jnl{Mem.~Soc.~Astron.~Italiana}}
\def\nphysa{\aaref@jnl{Nucl.~Phys.~A}}   
\def\physrep{\aaref@jnl{Phys.~Rep.}}   
\def\physscr{\aaref@jnl{Phys.~Scr}}   
\def\planss{\aaref@jnl{Planet.~Space~Sci.}}   
\def\procspie{\aaref@jnl{Proc.~SPIE}}   
\def\nar{\aaref@jnl{NewAR}}   
\def\icarus{\aaref@jnl{Icarus}}

\usepackage{natbib}

\hypersetup{
     colorlinks   = true,
     citecolor     = blue
} 

\newcommand{\mos}{\,ms$^{-1}$}
\newcommand{\kms}{\,kms$^{-1}$}

\newcommand\msini{\ifmmode{{\mathrm M} \sin i}\else${{\mathrm M} \sin i}$\fi}

\slugcomment{Accepted for publication in the Astrophysical Journal}

\shorttitle{Spin-orbit alignment of HATS-3}
\shortauthors{Addison et al.}

\begin{document}

\title{A Spin-Orbit Alignment for the Hot Jupiter HATS-3b\altaffilmark{$^{\dagger}$}}

\author{B. C. Addison \altaffilmark{1,2}, C. G. Tinney\altaffilmark{1,2}, D. J. Wright\altaffilmark{1,2}, D. Bayliss\altaffilmark{3}}

\email{b.addison@unsw.edu.au}

\altaffiltext{1}{Exoplanetary Science Group, School of Physics, University of New South Wales, Sydney, NSW 2052, Australia}
\altaffiltext{2}{Australian Centre of Astrobiology, University of New South Wales, Sydney, NSW 2052, Australia}
\altaffiltext{3}{Research School of Astronomy and Astrophysics, Australian National University, Canberra, ACT 2611, Australia}
\altaffiltext{$\dagger$}{Based on observations obtained at the Anglo-Australian Telescope, Siding Spring, Australia.}


\begin{abstract}
We have measured the alignment between the orbit of HATS-3b (a recently discovered, slightly inflated Hot Jupiter) and the spin-axis of its host star. Data were obtained using the CYCLOPS2 optical-fiber bundle and its simultaneous calibration system feeding the UCLES spectrograph on the Anglo-Australian Telescope. The sky-projected spin-orbit angle of $\lambda = 3\pm25^{\circ}$ was determined from spectroscopic measurements of Rossiter-McLaughlin effect. This is the first exoplanet discovered through the HATSouth transit survey to have its spin-orbit angle measured. Our results indicate that the orbital plane of HATS-3b is consistent with being aligned to the spin axis of its host star. The low obliquity of the HATS-3 system, which has a relatively hot mid F-type host star, agrees with the general trend observed for Hot Jupiter host stars with effective temperatures $>6250$~K to have randomly distributed spin-orbit angles.
\end{abstract}


\section{INTRODUCTION}
Recent measurements of the sky-projected spin-orbit angles\footnote{For clarity, we note that when we use the phrase `spin-orbit angle or obliquity', we are referring to the angle between the spin angular momentum vector of the host star and the orbital angular momentum vector of the planet.} for extra-solar planets (exoplanets) are revealing many surprises. These include planets on highly misaligned \citep[e.g.,][]{2012ApJ...757...18A}, polar \citep[e.g.,][]{2013ApJ...774L...9A}, and even retrograde orbits \citep[e.g.,][]{2010ApJ...722L.224B}. The dominant core-accretion model of planetary formation predicts that planets should form on nearly coplanar orbits with respect to their host star's equator \citep[e.g.,][]{1996Icar..124...62P,2011MNRAS.413L..71W,2014MNRAS.438L..31G} as is the case for the Solar System \citep{1993ARA&A..31..129L,2005ApJ...621L.153B}. Moreover, plausible models for orbital migration \citep[e.g., type I/II migration,][]{1996Natur.380..606L,2004A&A...417L..25A,2006ApJ...652L.133C} are expected to maintain this alignment. 

The total number of confirmed planets has ballooned to over 1700\footnote{\url{http://exoplanet.eu} \citep[or see,][]{2011A&A...532A..79S}, as of 2014 May 23. An alternative count can be obtained from the other main online exoplanet database, \url{http://exoplanets.org/} \citep[or see,][]{2011PASP..123..412W}, which yields a total number of confirmed planets of 1491 as of 2014 May 25. The difference between these two values is likely due to small differences in the degree of certainty required by those running the databases for a claimed exoplanet discovery to be considered `confirmed'.} including the 715 recently confirmed Kepler planets \citep{2014ApJ...784...45R}. This rapid growth in the number of known exoplanetary systems will allow the finer details of planet formation to finally be better understood. Measurements of the obliquity of these systems will play a critical role in this endeavor, providing key empirical evidence that will elucidate the complex formation and orbital evolution mechanisms of extra-solar planets.

Ground-based transit surveys, such as the Wide Angle Search for Planets \citep[WASP,][]{2006PASP..118.1407P}, the Hungarian Automated Telescope Network \citep[HATNet,][]{2004PASP..116..266B}, and its southern hemisphere counterpart HATSouth \citep{2013PASP..125..154B}, along with space-based transit surveys like CoRoT \citep{2008A&A...482L..17B} and Kepler \citep{2010Sci...327..977B,2013ApJS..204...24B}, are playing an important role in discovering and characterizing exoplanets. Measurements of spin-orbit alignments are increasingly becoming a critical part of the follow-up exoplanet characterization programs of the major exoplanet surveys. 

The number of systems with spin-orbit alignment measurements has dramatically increased over the past few years \citep[see e.g.,][]{2012ApJ...757...18A}. To date, seventy-six\footnote{\label{RM_foot}See Holt-Rossiter-McLaughlin Encyclopaedia (compiled by Ren\'{e} Heller and last updated 2014 June 05); \url{http://www.physics.mcmaster.ca/~rheller/index.html}} planetary systems have measured obliquities, and of these 31 show substantial misalignments ($>22.5^{\circ}$). The vast majority of the sampled planets, however, are on short-period orbits with Jupiter-like masses (i.e. Hot Jupiters). This is because the amplitude of the radial velocity anomaly used to determine the spin-orbit alignments is dependent on the square of the planet to star radius ratio, $\left(R_{P}/R_{\star}\right)^{2}$, and the rotational velocity of the host star ($v\sin i_{\star}$). Therefore, relatively few sub-Jovian, long-period, or multiple planet systems, have been studied and this parameter space remains largely unexplored. The large number of newly discovered Kepler planets has expanded the samples from which this parameter space can be explored\footnote{We note, however, that the majority of planets discovered from Kepler are orbiting stars too faint and/or host planets too small for follow-up spin-orbit alignment measurements.}, in particular via: obliquity measurements of Neptune-size planets \citep[e.g., a nearly polar orbit for the Super-Neptune, HAT-P-11b,][]{2010ApJ...723L.223W}; long-period planets \citep[e.g., stellar obliquity of the long-period planetary system HAT-P-17,][]{2013ApJ...772...80F}; and multiple planet systems \citep[see e.g.,][]{2012Natur.487..449S,2013ApJ...771...11A,2013Sci...342..331H,2014ApJ...783....9H}.

The majority of spin-orbit alignments have been determined using spectroscopic measurements of the Rossiter-McLaughlin effect. The Rossiter-McLaughlin effect was first predicted for eclipsing binary stars over one-hundred years ago by \citet{1893A&A...12...646H}, however, it was not observed until 1924 for eclipsing binary stars \citep{1924ApJ....60...15R,1924ApJ....60...22M} and 2000 was it observed for planets \citep{2000A&A...359L..13Q}. This effect is caused by the modification of the stellar spectrum as a transiting planet partially obscures the stellar disk of its host star, causing a radial velocity anomaly due to asymmetric distortions in the rotationally broadened stellar line profiles \citep{2005ApJ...622.1118O}. 

Three additional techniques include: planetary starspot-crossings \citep[e.g., the spin-orbit misalignment of HAT-P-11 was detected as a direct result of measured spot-crossing anomalies;][]{2011ApJ...743...61S}, asteroseismology (e.g., inclination of the multi-planet hosting star HR 8799 from asteroseismology; \citealt{2011ApJ...728L..20W}, the large obliquity for the multiplanet system Kepler-56 from asteroseismology measurements; \citealt{2013Sci...342..331H}), and gravity darkening \citep[e.g., orbital obliquity of KOI368.01 from gravity darkening;][]{2013ApJ...776L..35Z,2014ApJ...786..131A}.

In this paper, we present spectroscopic measurements of the Rossiter-McLaughlin effect for HATS-3 showing that the system is in spin-orbit alignment. HATSouth is a global network of wide-field telescopes capable of continuous 24~hr monitoring of specific regions in the sky \citep{2013PASP..125..154B}. To date, HATSouth has discovered five transiting extra-solar planets (HATS-1b, \citealt{2013AJ....145....5P};  HATS-2b, \citealt{2013A&A...558A..55M}; HATS-3b, \citealt{2013AJ....146..113B}; HATS-4b, \citealt{2014arXiv1402.6546J}; HATS-5b, \citealt{2014arXiv1401.1582Z}). The obliquity for HATS-3b, reported here, is the first that was measured for a system discovered by the HATSouth planet search.

\section{OBSERVATIONS}
HATS-3 is a mid-late F star with a mass of $M_{\star}=1.209\pm0.036$\,$M_{\odot}$, radius of $R_{\star}=1.404\pm0.030$\,$R_{\odot}$, effective temperature of $T_{eff}=6351\pm76$\,K, age of $3.2^{+0.6}_{-0.4}$\,Gyr, $453\pm11$\,pc\ away from Earth, and has moderate rotation ($v \sin i_{\star}=9.12\pm1.31$\,\kms) as reported by \citet{2013AJ....146..113B}. The planet is a Hot Jupiter with a mass of $M_{P}=1.071\pm0.136$\,$M_{J}$, slightly inflated with a radius of $R_{P}=1.381\pm0.035$\,$R_{J}$, orbiting at distance of $a=0.0485^{+0.0004}_{-0.0006}$\,AU, and an orbital period of $P=3.547851\pm0.000005$ \citep{2013AJ....146..113B}. HATS-3b is the third planet discovered from HATSouth survey and a good candidate to have a measurable Rossiter-McLaughlin effect anomaly due to a high $v \sin i_{\star}$ and large $R_{P}$ \citep{2013AJ....146..113B}.

\subsection{Simultaneous Photometric Observations on the FTS 2m}
Previous studies have shown the value of acquiring simultaneous photometric observations of exoplanetary systems for which Rossiter-McLaughlin effect radial velocity measurements are being obtained \citep[e.g.,][]{2011ApJ...743...61S,2013A&A...549A..35O,2013ApJ...775...54S}. Such observations are particularly useful in disentangling the signature of starspot-crossings from true Rossiter-McLaughlin induced radial velocity variations \citep[e.g., Kepler-63b,][]{2013ApJ...775...54S}. Starspot-crossings have been observed in both radial velocity and transit photometry data of other transiting planetary systems (e.g., radial velocity and photometric spot-crossing anomalies of Kepler-63b, \citealt{2013ApJ...775...54S}; and photometric spot-crossing anomalies of HAT-P-11, \citealt{2011ApJ...743...61S}). They can introduce systematic radial velocity anomalies that are comparable in size to, and superimposed with, the Rossiter-McLaughlin effect, making an accurate determination of $\lambda$ more difficult. Spot-crossings are, however, easily observable with photometry for large $R_{P}$ and large spots. Simultaneous photometry can, therefore, be used to effectively remove spot-crossing anomalies from radial velocity measurements of the Rossiter-McLaughlin effect \citep[e.g., Kepler-63b,][]{2013ApJ...775...54S}. An additional benefit of photometry of spot-crossing events is that they can provide constraints on the true orbital obliquity \citep[instead of just the projected obliquity, as described in][]{2011ApJ...740L..10N} of planets that transit spotty stars \citep[e.g., HATS-2b,][]{2013A&A...558A..55M}.

\begin{figure}
\centering
\includegraphics[width=1.0\linewidth]{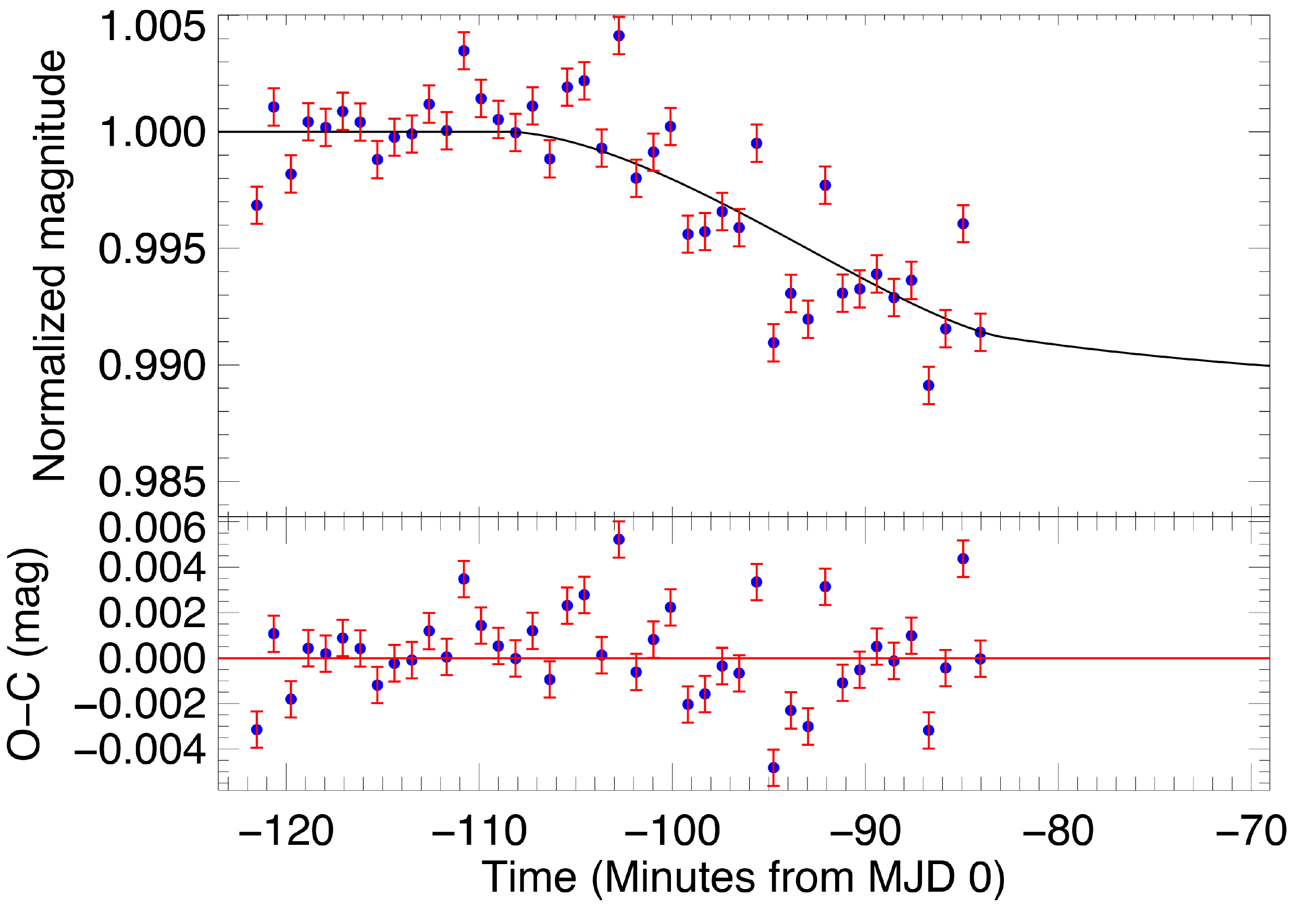}
\caption[LoF entry]{Simultaneous transit photometry for HATS-3b obtained on 2013 August 27 starting $\sim 15$~mins before transit ingress and ending $\sim 20$~mins into the transit (near second contact). A best fit model has been over-plotted. Residuals are shown on the bottom plot. The transit photometry was used to constrain the mid transit time.
\\ [+1.5ex]
(A color version of this figure will be available in the online journal.)}
\label{Figure:transit}
\end{figure}

We obtained simultaneous photometric transit observations of the HATS-3 transit on the night of 2013 August 27 using the Faulkes Telescope South (FTS) to provide an updated constraint on the time of transit ingress. The FTS is located at the same site as the AAT (Siding Spring Observatory) and is part of the Las Cumbres Global Telescope (LCOGT) Network. Observations were obtained using the ``Spectral'' imaging camera in 2\,$\times$\,2 binned readout mode. The telescope was moderately defocused to avoid saturating on longer exposures and to minimize flat-fielding errors. The $i$-band filter and 30\,s integration times (cadence of 50\,s including the 20\,s readout) were used for our transit observation. The data was reduced to calibrated fits files using the LCOGT data reduction pipeline. Aperture photometry was then performed using Source Extractor \citep{1996A&AS..117..393B}. Unblended, non-variable reference stars from the images were selected to de-trend the photometry, although this was the limiting factor in the precision of the photometry as none of the reference stars were as bright as HATS-3 in the 10\textquotesingle\,$\times$\,10\textquotesingle\ field of view. Additionally the star passed near zenith during the observations, resulting in a very large systematic error in the photometry due to rapid pupil rotation. We did not attempt to derive photometry during this period of time. The final result was that we obtained simultaneous photometric follow-up for HATS-3 for a total of 38\,min on the 2013 August 27 starting $\sim 15$\,min before transit ingress. The lightcurve is presented in Figure~\ref{Figure:transit} and Table~\ref{table:Photometry}.

\begin{table*}
\centering
\begin{minipage}{140mm}
\caption{Differential Photometry of the 2013 August 27 HATS-3b transit.}
\centering
\begin{tabular}{l c c c l c c}
\hline\hline \\ [-2.0ex]
Time & $\mathrm{R_{Mag}}$ & $\mathrm{\sigma_{R_{Mag}}}$ &     & Time & $\mathrm{R_{Mag}}$ & $\mathrm{\sigma_{R_{Mag}}}$ \\
BJD-2400000 &  &  &     	& BJD-2400000 &  & \\ [0.5ex]
\hline \\ [-2.0ex]
56531.95360 & -0.00315 & 0.00080 &     \hspace{10 mm}		& 56531.96725 & -0.00199 & 0.00080 \\
56531.95421 &  0.00107 & 0.00080 &     \hspace{10 mm}		& 56531.96787 & -0.00087 & 0.00080 \\
56531.95483 & -0.00181 & 0.00080 &     \hspace{10 mm}		& 56531.96848 &  0.00023 & 0.00080 \\
56531.95545 &  0.00043 & 0.00080 &     \hspace{10 mm}		& 56531.96911 & -0.00439 & 0.00080 \\
56531.95608 &  0.00019 & 0.00080 &     \hspace{10 mm}		& 56531.96972 & -0.00428 & 0.00080 \\
56531.95669 &  0.00088 & 0.00080 &     \hspace{10 mm}		& 56531.97034 & -0.00342 & 0.00080 \\
56531.95732 &  0.00042 & 0.00080 &     \hspace{10 mm}		& 56531.97095 & -0.00411 & 0.00080 \\
56531.95794 & -0.00119 & 0.00080 &     \hspace{10 mm}		& 56531.97158 & -0.00049 & 0.00080 \\
56531.95855 & -0.00023 & 0.00080 &     \hspace{10 mm}		& 56531.97219 & -0.00905 & 0.00080 \\
56531.95918 & -0.00009 & 0.00080 &     \hspace{10 mm}		& 56531.97281 & -0.00693 & 0.00080 \\
56531.95980 &  0.00119 & 0.00080 &     \hspace{10 mm}		& 56531.97343 & -0.00804 & 0.00080 \\
56531.96043 &  0.00005 & 0.00080 &     \hspace{10 mm}		& 56531.97405 & -0.00229 & 0.00080 \\
56531.96105 &  0.00348 & 0.00080 &     \hspace{10 mm}		& 56531.97467 & -0.00692 & 0.00080 \\
56531.96167 &  0.00143 & 0.00080 &     \hspace{10 mm}		& 56531.97529 & -0.00674 & 0.00080 \\
56531.96229 &  0.00053 & 0.00080 &     \hspace{10 mm} 		& 56531.97590 & -0.00610 & 0.00080 \\
56531.96291 & -0.00003 & 0.00080 &     \hspace{10 mm}		& 56531.97652 & -0.00711 & 0.00080 \\
56531.96352 &  0.00111 & 0.00080 &     \hspace{10 mm}		& 56531.97715 & -0.00637 & 0.00080 \\
56531.96414 & -0.00116 & 0.00080 &     \hspace{10 mm}		& 56531.97777 & -0.01088 & 0.00080 \\
56531.96477 &  0.00192 & 0.00080 &     \hspace{10 mm}		& 56531.97838 & -0.00845 & 0.00080 \\
56531.96538 &  0.00219 & 0.00080 &     \hspace{10 mm}		& 56531.97901 & -0.00394 & 0.00080 \\
56531.96601 & -0.00070 & 0.00080 &     \hspace{10 mm}		& 56531.97963 & -0.00860 & 0.00080 \\
56531.96663 &  0.00413 & 0.00080 &	   \hspace{10 mm}		& \\
\hline \\ [-2.5ex]
\end{tabular}
\label{table:Photometry}
\end{minipage}
\end{table*}

\subsection{Spectroscopic Observations with CYCLOPS2}
\begin{table}
\centering
\begin{minipage}{82mm}
\caption{Summary of HATS-3b Transit Spectroscopic Observations.}
\centering
\resizebox{\columnwidth}{!}{%
\begin{tabular}{l c c}
\hline\hline \\ [-2.0ex]
 & 2013 August 20 & 2013 August 27 \\ [0.5ex]
\hline \\ [-2.0ex]
UT Time of Obs & 08:40-14:32\,UT & 09:02-16:50\,UT \\
Cadence & 1120\,s & 1375-1675\,s \\
Readout Times & 120\,s & 175\,s \\
Readout Speed & Fast & Normal \\
Readout Noise & 5.35\,$e^{-}$ & 3.19\,$e^{-}$ \\
S/N (/2.5~pix at $\lambda=5490$~\AA{}) & 3-5 & 14-19 \\
Resolution (${\lambda}/{\Delta}{\lambda}$) & 70,000 & 70,000 \\
Number of Spectra & 20 & 19 \\
Seeing &  $1.5^{\texttt{"}}$ & $0.7^{\texttt{"}}$-$1.1^{\texttt{"}}$ \\
Weather Conditions & Some cirrus & Clear \\
Airmass Range & 1.0-2.3 & 1.0-1.8 \\
\hline \\ [-2.5ex]
\end{tabular}}
\label{table:spectro_obs}
\end{minipage}
\end{table}

The spectroscopic observations of HATS-3b were carried out using the CYCLOPS2 instrument on the AAT. Providing full details of the CYCLOPS2 instrument design and specifications are beyond the scope of this work, and we direct the interested reader to \citet{2012SPIE.8446E..3AH} for that information. The instrumental set-up and observing strategy for HATS-3b transit observations substantially followed that employed by \citet{2013ApJ...774L...9A}. The observations were calibrated using both a Thorium-Argon calibration lamp (ThAr), to illuminate all on-sky fibers, and a ThXe lamp to illuminate the simultaneous calibration fiber.


We observed a transit of HATS-3b on the night of 2013 August 20, starting $\sim20$\,min before ingress and finishing $\sim2$\,hr after egress. The observations are summarized in Table~\ref{table:spectro_obs}. A total of 20 spectra were obtained on that night (twelve during the $\sim3.5$\,hr transit) in reasonable observing conditions (seeing around $1.5^{\texttt{"}}$ with some high-level cirrus). The airmass at which HATS-3 was observed ranged from 2.3 at the start of the night, 1.2 near mid-transit, close to 1.0 at egress, and 1.1 at the end of the observations. We obtained a $\mathrm{S/N}=18$ per 2.5 pixel resolution element at $\lambda=5490$\,\AA{} (in total over all 16 fibers) at an airmass of 1.0 and in $1.5^{\texttt{"}}$ seeing. HATS-3 on this night was at a distance of $\sim15^{\circ}$ from the full Moon. However, cross-correlation of our data with a solar-like spectral mask demonstrated no obvious signatures of solar spectrum contamination near the main cross-correlation peak and we conclude that the observations were not significantly impacted by lunar contamination.

A second transit was observed on the night of 2013 August 27, starting $\sim2$\,hr before ingress and finishing $\sim2$\,hr after egress (see Table~\ref{table:spectro_obs}). A total of 19 spectra were acquired -- including nine during the $\sim3.5$\,hr transit. On this night, observations were obtained in a slower ``NORMAL'' readout mode. A quick-look analysis of the August 20 data had indicated that the additional read-noise delivered by the FAST readout mode was not worth the improved cadence delivered by the shorter read-time. The observing conditions were excellent for Siding Spring Observatory with seeing between $0.7^{\texttt{"}}$ to $1.1^{\texttt{"}}$ and clear skies for the whole night. The airmass at which HATS-3 was observed that night varied between 1.4 for the first exposure, $\sim1.1$ near mid transit, and 1.8 for the last exposure. We obtained a $\mathrm{S/N}=28$ per 2.5 pixel resolution element at $\lambda=5490$\,\AA{} (in total over all 16 fibers) on this night when the star was observed at an airmass of 1.0 with $0.7^{\texttt{"}}$ seeing.

\section{Analysis}\label{sec:RM_effect}
We used the Exoplanetary Orbital Simulation and Analysis Model \citep[ExOSAM,][]{2013ApJ...774L...9A} to determine the best fit parameters from both transit photometry and Rossiter-McLaughlin effect measurements.

\subsection{Transit Modeling}
The ExOSAM lightcurve analysis model is a significantly improved version of an earlier model called Exopanetary Pixelization Transit Model \citep[see][]{2010JSARA...3...45A}. ExOSAM utilizes the small planet approximation of \citet{2002ApJ...580L.171M}, which assumes that the surface brightness of the star directly underneath the disk of the planet is constant (an excellent approximation for the vast majority of transiting systems, including HATS-3b), to determine the limb-darkened stellar flux blocked by the planet. Limb-darkening is described in our model using either a linear or quadratic law. For HATS-3b, a quadratic limb-darkening law (see equations\,\ref{equat:Io} \& \ref{equat:Fblocked}) was used.

\begin{equation}\label{equat:Io}
I_{0} = \frac{6}{\pi R_{\star}^{2} \left(6 - 2q_{1} - q_{2}\right)}
\end{equation}

\begin{multline}\label{equat:Fblocked}
\mathrm{F_{b}} = I_{0}A_{P}\left[1-q_{1}\left\{1-\sqrt{\left|1-\frac{d_{c}^{2}}{R_{\star}^{2}}\right|}\right\}\right] \\
-I_{0}A_{P}\left[q_{2} \left\{1 - \sqrt{\left|1-\frac{d_{c}^{2}}{R_{\star}^{2}}\right|}\right\}^{2}\right]
\end{multline}

Where $I_{0}$ is the central intensity of the stellar surface such that $F_{b}$ is normalized to one \citep[see appendix A of][for simple derivation of $I_{0}$]{2010ApJ...709..458H}, $R_{\star}$ is the stellar radius (in m), and $q_{1}$ and  $q_{2}$ are the limb-darkening coefficients. $F_{b}$ is the total fractional (normalized) intensity blocked by the planet, $A_{P}$ is the area of the planet in front of the stellar disk, and $d_{c}$ is the apparent distance between the center of the planetary disk to the center of the stellar disk along the transit chord as viewed from the Earth. $d_{c}$ in equation\,\ref{equat:Fblocked} determines the uniform limb-darkened surface brightness of the star directly underneath the disk of the transiting planet.

ExOSAM uses a total of 10 input parameters to calculate the best-fit transit lightcurves. Of these, ExOSAM can solve for eight, namely: the planet-to-star radius ratio ($R_{p}/R_{\star}$); the orbital inclination angle ($i$); the orbital period ($P$); the orbital eccentricity ($e$); argument of periastron ($\varpi$); the two coefficients ($q_{1}$ \&  $q_{2}$) in the quadratic limb-darkening equation; and the mid-transit time ($T_{0}$). The final two parameters, the stellar mass ($M_{\star}$ estimated from spectroscopy) and the planet mass ($M_{p}$ estimated from radial velocity data), are held fixed. The eight free parameters are derived using a well-sampled grid search and minimizing $\chi^{2}$ between the observed transit photometry and modeled lightcurve. The $1\sigma$ confidence levels in the free parameters are determined through the $\Delta \chi^{2}$ method \citep{press1992numerical} which is based on the normal probability distribution of $\chi^{2}$ as a function of the confidence level and degrees of freedom.

\subsection{Rossiter-McLaughlin Effect Modeling}
The ExOSAM Rossiter-McLaughlin analysis model described in \citet{2013ApJ...774L...9A} instead uses 16 input parameters, of which we fix 14 in order to allow us to accurately determine the projected orbital obliquities $\lambda$ and projected stellar rotational velocities $v\sin i_{\star}$ by fitting radial velocity data taken during a transit event (when the Rossiter-McLaughlin effect is observable). The 14 fixed parameters are as follows: the planet-to-star radius ratio ($R_{p}/R_{\star}$); the orbital inclination angle ($i$); the orbital period ($P$); the mid-transit time ($T_{0}$) at the epoch of observation; the radial velocity offset ($V_{d}$) between our data sets and published data sets; a velocity offset term ($V_{s}$) accounting for systematic effects between our data sets taken over multiple nights; planet-to-star mass ratio ($M_{p}/M_{\star}$); orbital eccentricity ($e$); argument of periastron ($\varpi$); two adopted quadratic limb-darkening coefficients ($q_{1}$ and $q_{2}$); the micro-turbulence velocity ($\xi_{t}$); the macro-turbulence velocity ($v_{mac}$); and the center-of-mass velocity ($V_{T_{P}}$) at published epoch $T_{P}$.

ExOSAM models both the radial velocity from the motion of the host star due to the orbiting planet and the velocity anomaly due to the Rossiter-McLaughlin effect, using the analytical approach of \citet{2010ApJ...709..458H} as described in \citet{2013ApJ...774L...9A}. We have included a Monte Carlo simulation in our model to obtain more robust estimates of the uncertainties on the spin-orbit angle ($\lambda$) and the rotational velocity ($v\sin i_{\star}$) of the host star from the given uncertainties on other fixed input parameters. Confidence intervals for $\lambda$ and $v\sin i_{\star}$ ($\Delta \lambda$, $\Delta v\sin i_{\star}$) are derived from Monte Carlo simulations as adopted from \citet{press1992numerical} and given by equations\,\ref{equat:lambda_uncert} \& \ref{equat:vsini_uncert}: 

\begin{equation}\label{equat:lambda_uncert}
\Delta \lambda=\sqrt{\frac{1}{N_{MC}-1}*\sum_{i=1}^{N_{MC}}\left[\left(\lambda_{o}-\lambda_{i}\right)^{2}+\Delta\lambda_{i}^{2}\right]}
\end{equation}

\begin{multline}\label{equat:vsini_uncert}
\Delta v\sin i=\sqrt{\frac{1}{N_{MC}-1}}\times \\
\sqrt{\sum_{i=1}^{N_{MC}}\left[\left(v\sin i_{(o)}-v\sin i_{(i)}\right)^{2}+\Delta v\sin i_{(i)}^{2}\right]}
\end{multline}

where $N_{MC}$ is the number of Monte Carlo simulations, $\lambda_{o}$ and $v\sin i_{(o)}$ are the best overall $\lambda$ and $v\sin i_{\star}$ from all Monte Carlo runs (as determined from the minimum $\chi^{2}$), $\lambda_{i}$ and $v\sin i_{(i)}$ are the best $\lambda$ and $v\sin i_{\star}$ from the $i^{th}$ Monte Carlo run, and $\Delta\lambda_{i}$ and $\Delta v\sin i_{(i)}$ are the $1\sigma$ confidence levels of $\lambda$ and $v\sin i_{\star}$ determined through the $\Delta \chi^{2}$ method for the $i^{th}$ Monte Carlo run.

\subsection{Transit Analysis}
We modeled the partial transit lightcurve of HATS-3 using the ExOSAM model. The only parameter we solve for from this data is the mid-transit time ($T_{0}$) at the epoch of observation, which we wish to use as a fixed parameter in the subsequent Rossiter-McLaughlin effect modeling. We fixed the nine other parameters to the values published in \citet{2013AJ....146..113B} and used the quadratic limb-darkening law (see equations\,\ref{equat:Io} \& \ref{equat:Fblocked}) with the published Sloan r-filter limb-darkening coefficients from \citet{2013AJ....146..113B}, $q_{1}=0.2592$ and $q_{2}=0.3725$, as fixed inputs in our modeled lightcurve.

The best-fitting value for $T_{0}$ and its $1\sigma$ confidence level are derived using a well-sampled grid search that minimizes $\chi^{2}$ between the observed transit photometry and modeled lightcurve. The step size used in the grid search was 2\,s and the range searched was barycentric Julian dates between 2456532.01455\,d to 2456532.06455\,d. The predicted mid transit time on 2013 August 27 was $2456532.03955\pm0.00014$\,d and was determined from the published orbital period $P$ and mid-transit time of observation in \citet{2013AJ....146..113B}. The mid transit time we determined from our photometry is $2456532.03799\pm0.00028$\,d. 

\subsection{Rossiter-McLaughlin Analysis}
The spectroscopic data were reduced using custom MATLAB routines, which trace each fiber and optimally extract each spectral order as outlined previously in \citet{2013ApJ...774L...9A}. Each of the 16 fibers, in each of the 18 useful orders, is used to estimate a radial velocity (and associated uncertainty) by cross-correlation with a synthetic spectrum of similar spectral type using the IRAF\footnote{IRAF is distributed by the National Optical Astronomy Observatories, which are operated by the Association of Universities for Research in Astronomy, Inc., under cooperative agreement with the National Science Foundation \citep{1986SPIE..627..733T}.} task, \textit{fxcor}. We created the synthetic template spectrum of a mid-F star ($T_{eff_{\star}}=6500$~K and log~$g_{\star}=4.0$) using SYNSPEC\footnote{Information on SYNSPEC can be found at \url{http://nova.astro.umd.edu/Synspec49/synspec.html} and briefly described in the following publication \citet{1985BAICz..36..214H}.}, a general spectrum synthesis program. \textit{Fxcor} uses the standard cross-correlation technique developed by \citet{1979AJ.....84.1511T}. We observed several radial velocity standard stars, including HD 206395, HD 10700, and HD 6735. We carried out tests using both these radial velocity standard observations and synthetic spectra as cross-correlation templates, and found that the lowest inter-fider\footnote{We use the term `fider' to refer to the spectrum extracted from a single fiber in a single spectral order in the echellogram.} velocity scatter was obtained using the spectrum of the synthetic template, and we therefore adopted this for use in the subsequent analysis. The weighted average velocities for each observation were computed using the method described in \citet{2013ApJ...774L...9A} and the uncertainties for each weighted velocity were estimated from the weighted standard deviation of the fider velocity scatter. Our weighted radial velocities for the two transit observations, including their uncertainties and total S/N, are shown in Table\,\ref{table:RVs}.

$\lambda$ and $v\sin i_{\star}$ were determined from the best-fit model of the Rossiter-McLaughlin effect using ExOSAM. The mid-transit time $T_{0}$ was fixed to the best derived value from our simultaneous photometry. The velocity offset, $V_{s}$, between our data sets taken on August 20 and 27 was determined by finding the different $V_{d}$ offsets between the Bayliss et al. data set and our data sets for each night separately (August 20 $V_{d_{1}}$ and August 27 $V_{d_{2}}$) and applying the difference to our combined data set of both nights ($V_{s}=V_{d_{1}}-V_{d_{2}}$). The other seven input parameters ($M_{p}/M_{\star}$, $e$, $\varpi$, $q_{1}$, $q_{2}$, $\xi_{t}$, and $V_{T_{P}}$) were adapted from \citet{2013AJ....146..113B}.

The confidence intervals for our $\lambda$ and $v\sin i_{\star}$ were derived from running 5000 Monte Carlo iterations. For each iteration, a synthetic data set was generated by drawing from a normal distribution for each radial velocity datum and its $1\sigma$ uncertainty. We also drew from randomly generated Gaussian distributions for the model parameters $R_{p}/R_{\star}$, $i$, $P$, $T_{0}$, and $V_{d}$ about their mean and standard deviation as given in \citet{2013AJ....146..113B}. We determined that the remaining model parameters $M_{p}/M_{\star}$, $e$, $\varpi$, $q_{1}$, $q_{2}$, $\xi_{t}$, and $V_{T_{P}}$ negligibly contribute to the overall uncertainty in $\lambda$ and $v\sin i_{\star}$ and so held these fixed in our simulations. The input parameters and their uncertainties as adopted are given in Table\,\ref{table:Parameters}.

Best-fitting values for $\lambda$ and $v\sin i_{\star}$ for each Monte Carlo run were determined from a grid search that minimized $\chi^{2}$ on a uniformly distributed, randomly generated set of 120 $\lambda$ values between $-60^{\circ} \leq \lambda \leq 60^{\circ}$ and 75 $v\sin i_{\star}$ values between $1.0 \leq v\sin i_{\star} \leq 12.0$\,\kms. The ranges for $\lambda$ and $v\sin i_{\star}$ were chosen based on a quick inspection of the observed Rossiter-McLaughlin effect velocity anomaly. We then determined the best overall $\lambda$ and $v\sin i_{\star}$ and these were used to compute their $1\sigma$ uncertainties ($\Delta \lambda$) and ($\Delta v\sin i_{\star}$) as given in equations\,\ref{equat:lambda_uncert} \& \ref{equat:vsini_uncert}. 

Table\,\ref{table:Parameters} shows the final best-fit parameters and their uncertainties for eccentricity fixed at zero. \citet{2013AJ....146..113B} computed two sets of planetary parameters for HATS-3b: one based on a fixed circular orbit and another allowing the eccentricity to float. When they allowed the eccentricity to vary, they determined that the best fit eccentricity was $e=0.25 \pm 0.10$. Unfortunately, the available radial velocity data does not decisively indicate whether the orbit is circular or eccentric. Given that the planet is in a $\sim 3.5$\,day orbit and the system is over 3\,Gyr old, we consider an $e\not\approx0$ as unlikely. Such a planet should have been tidally circularized long ago, thus our analysis is based on the assumption that $e=0$.

\begin{figure}
\centering
\includegraphics[width=1.0\linewidth]{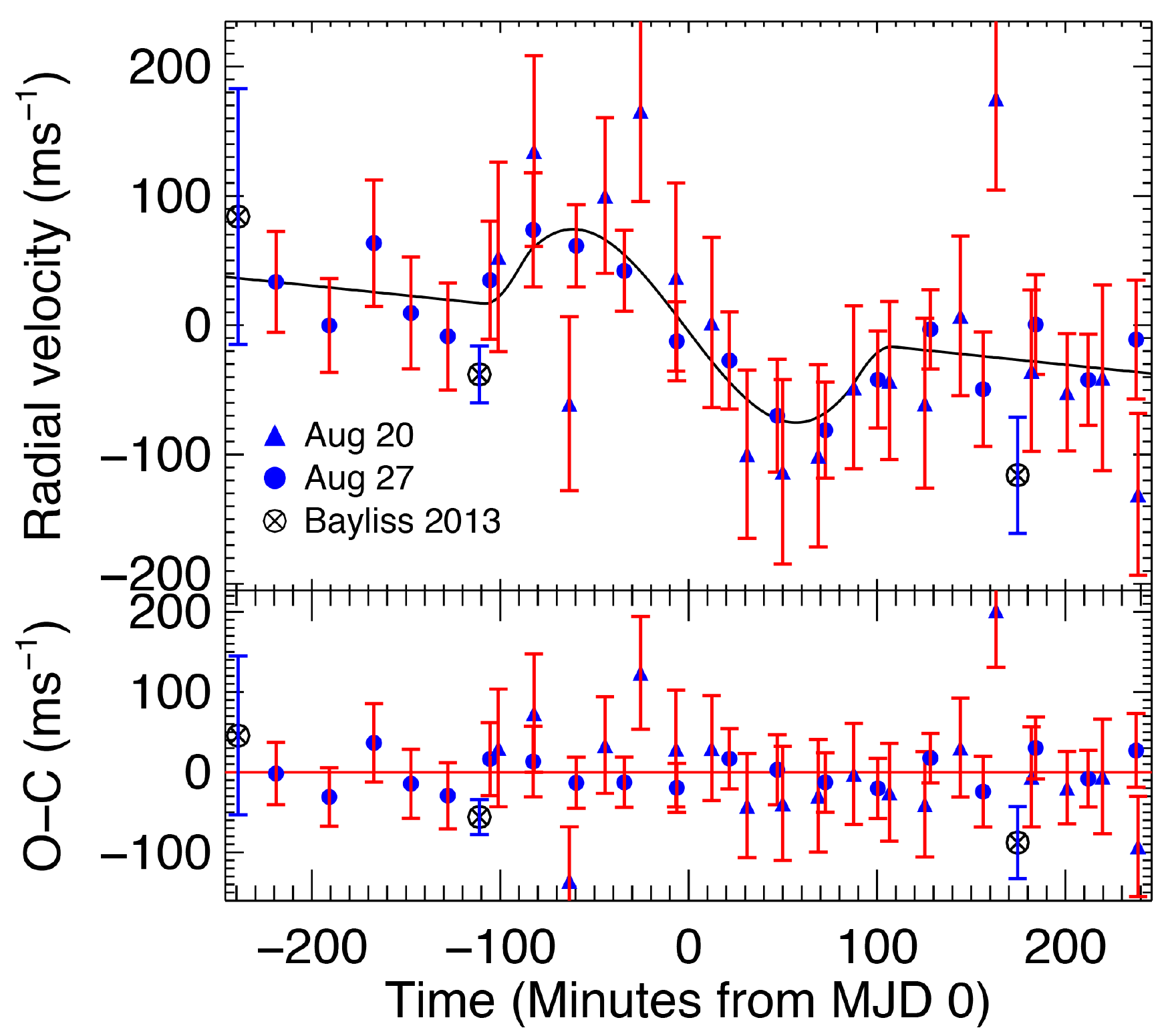}
\caption[LoF entry]{Spectroscopic radial velocities of two separate HATS-3b transits that have been phased-shifted to 0 Modified Julian Date (we set the mid transit to be at 0 MJD). Velocities from just before, during, and after the transit are plotted as a function of time (MJD) along with the best fitting model and corresponding residuals. The filled upward pointing blue triangles and the filled blue circles with red error bars are radial velocities we measured on 2013 August 20 and 27 respectively with our estimated uncertainties. The three black circles with an x and with blue error bars are previously published velocities by \citet{2013AJ....146..113B} using their quoted uncertainties. The zero velocity offsets for our August 20 and 27 datasets as well as the velocity offset between them were determined from the \citet{2013AJ....146..113B} radial velocities.
\\ [+1.5ex]
(A color version of this figure will be available in the online journal.)}
\label{Figure:RM}
\end{figure}

\begin{figure}
\centering
\includegraphics[width=1.0\linewidth]{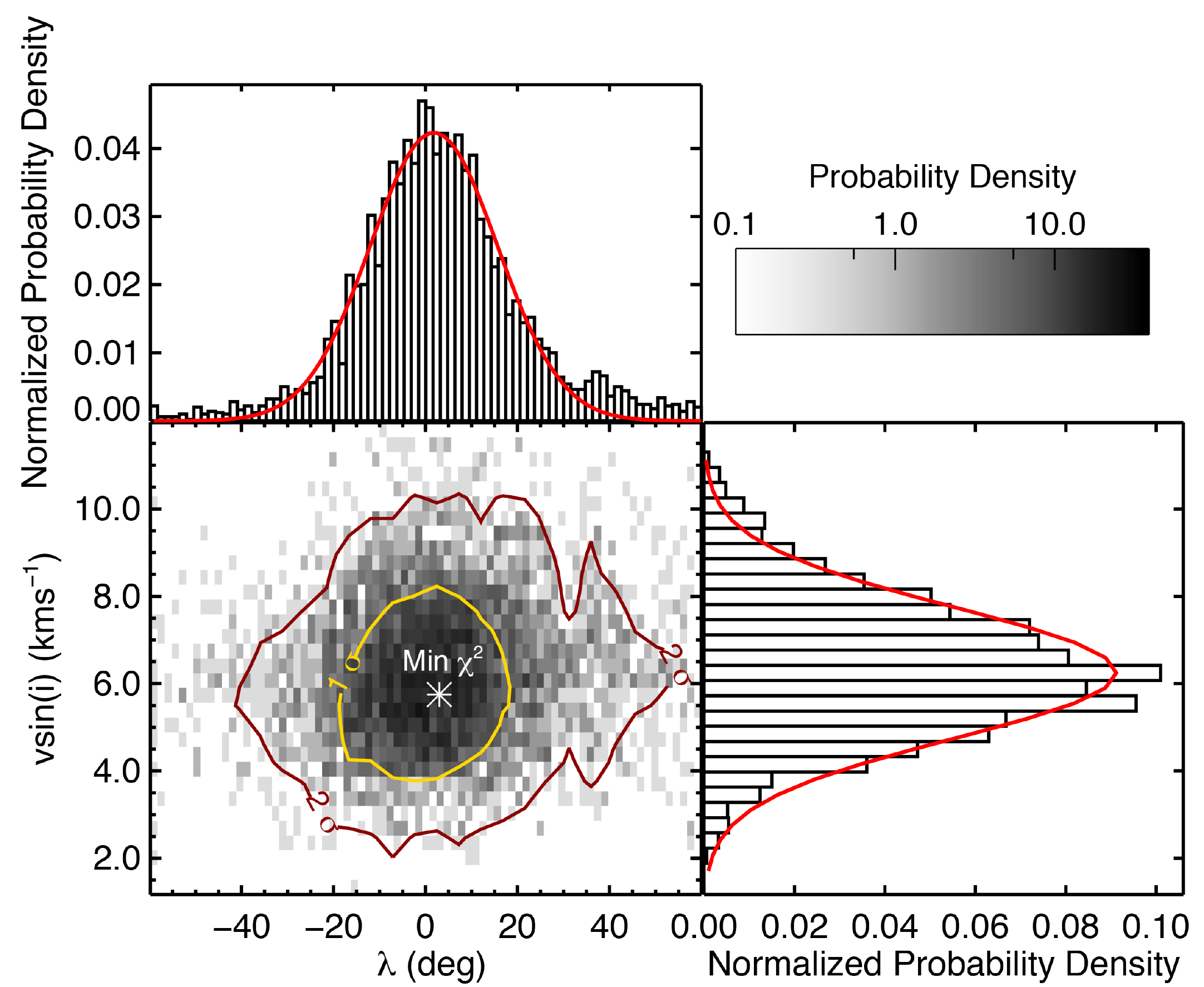}
\caption[LoF entry]{The density distribution of $\lambda$ and $v\sin i_{\star}$ from our Monte Carlo simulation. The contours show the 1 and 2 $\sigma$ confidence regions (in yellow and red respectively). Fitted Gaussians (in red) are shown separately for $\lambda$ (above) and $v\sin i_{\star}$ (right).
\\ [+1.5ex]
(A color version of this figure will be available in the online journal.)}
\label{Figure:distro}
\end{figure}

Figure\,\ref{Figure:RM} shows the modeled Rossiter-McLaughlin anomaly with the observed velocities over-plotted. The density distribution for $\lambda$ and $v\sin i_{\star}$ resulting from the Monte Carlo simulations is shown in Figure\,\ref{Figure:distro}, along with the location of the $\chi^{2}$ minimum as well as the $1\sigma$ and $2\sigma$ confidence contours. Normalized density functions collapsed into $\lambda$ and $v\sin i_{\star}$ are also shown, along with fitted Gaussians.

We clearly detect the Rossiter-McLaughlin effect from the combined data sets of August 20 and 27 as a positive anomaly between $\sim100$ minutes prior to mid-transit ($0$\,MJD) to $\sim0$\,MJD and then negative anomaly between $\sim0$\,MJD to $\sim100$ minutes after mid-transit. That is, the planet initially transits across the blue-shifted hemisphere during ingress, crosses the mid transit point near the stellar rotation axis, and then transits across the red-shifted hemisphere during egress. This produces a nearly symmetrical velocity anomaly as seen in Figure\,\ref{Figure:RM}. The lack of any asymmetry suggests that the planet is in an orbit well-aligned to the rotational axis of its host star (i.e. that is in ``spin-orbit alignment'').

For the August 20 and 27 datasets, we obtained the projected obliquity as $\lambda = 3^{\circ} \pm 25^{\circ}$ and stellar rotation velocity as $v\sin i_{\star} = 5.75 \pm 2.98$\,\kms. We also conducted the analysis on the datasets from the two nights separately and obtained $\lambda = -8^{\circ} \pm 55^{\circ}$ and $v\sin i_{\star} = 7.50 \pm 4.18$\,\kms\ for August 20; and $\lambda = 8^{\circ} \pm 33^{\circ}$ and $v\sin i_{\star} = 5.25 \pm 3.01$\,\kms\ for August 27. The two nights produce consistent results, though the August 20 data delivers significantly higher uncertainties for $\lambda$ and $v\sin i_{\star}$ due to the lower S/N spectra obtained on that night.

Examination of the normalized residuals to the model (R$_{Norm}$ as defined in equation\,\ref{equat:Rnorm}) and reduced $\chi^2$ ($\chi^{2}_{red}$) suggests that our estimated velocity uncertainties may have been overestimated, as we find $R_{Norm}=0.62$ and $\chi^{2}_{red}=0.90$. We therefore experimented with an empirical adjustment of those uncertainties in a manner similar to \citet{2004ApJ...600L..75B}. An updated solution with these adjusted uncertainties produces values for $\lambda$ and $v\sin i_{\star}$ consistent with those obtained previously, but with smaller uncertainties ($\lambda = 4^{\circ} \pm 16^{\circ}$ and $v\sin i_{\star} = 5.75 \pm 1.70$\,\kms). However, in the absence of a plausible cause for our uncertainties being overestimated, we favor our original values, but we do quote both solution uncertainties in Table\,\ref{table:Parameters}.

\begin{equation}\label{equat:Rnorm}
R_{Norm}=\frac{\sum_{i=1}^{N_{data}}\left[\frac{\left|O-C\right|_{i}}{\sigma_{i}}\right]}{N_{data}}
\end{equation}

We therefore checked our $v\sin i_{\star}$ estimation in two additional ways. First, we fitted a rotationally broadened Gaussian to the cross-correlation peak produced by each of the HATS-3 spectra taken that night (summed over all echelle orders) to obtain $v\sin i_{\star} = 5.2 \pm 0.6$\,\kms. Second, we fitted a rotationally broadened Gaussian to a least-squares deconvolution line profile for each spectral order (in a similar manner to that used in \citealt{2013ApJ...774L...9A}) of the three best spectra of HATS-3 taken on August 27, giving $v\sin i_{\star} = 5.3 \pm 0.7$\,\kms. Both of these estimates are consistent with the value determined from our Rossiter-McLaughlin fitting. Most critically, if the $v\sin i_{\star}$ in HATS-3 were as large as that presented in \citet{2013AJ....146..113B} of $9.12 \pm 1.31$\,\kms, we would have obtained a velocity anomaly $50\%$ larger than we actually observed. We are therefore confident in adopting a $v\sin i_{\star}=5.75 \pm 2.98$\,\kms\ for this system. One possible explanation for the discrepancy between the published $v\sin i_{\star}$ of \citet{2013AJ....146..113B} and our measured $v\sin i_{\star}$ could be from not accounting for stellar macroturbulence, which can contribute significantly to the overall absorption line broadening observed in stars. There exists a degeneracy between rotational broadening and macroturbulence broadening, as determined by the width of the stellar absorption lines, which makes disentangling each of their contributions to the overall broadening difficult to measure \citep{2014arXiv1402.6546J}.

\begin{table*}[ht]
\centering
\begin{threeparttable}[b]
\caption{Radial velocities for HATS-3 (fiber and order averaged) taken on 2013 August 20 \& 27.}
\centering
\begin{tabular}{l c c c c l c c c}
\hline\hline \\ [-2.0ex]
Time & RV & S/N at & In/Out &  & Time & RV & S/N at & In/Out \\
BJD-2400000 & (\mos) & $\lambda$=5490\AA{} & Transit &  & BJD-2400000 & (\mos) & $\lambda$=5490\AA{} & Transit \\ [0.5ex]
\hline \\ [-2.0ex]
56524.85790\tnote{a,b} & -40317 $\pm$ 103 & 18.6 & Out &     \hspace{10 mm}		& 56525.20517\tnote{b} & -40763 $\pm$ 82 & 25.9 & Out \\
56524.87216\tnote{a} & -40741 $\pm$ 73 & 16.6 & Out &     \hspace{10 mm}		& 56531.88589 & -40761 $\pm$ 39 & 26.6 & Out \\
56524.88526\tnote{a} & -40659 $\pm$ 74 & 16.3 & In &     \hspace{10 mm}		& 56531.90552 & -40794 $\pm$ 36 & 29.3 & Out \\
56524.89835\tnote{a} & -40855 $\pm$ 67 & 16.7 & In &     \hspace{10 mm}		& 56531.92201 & -40731 $\pm$ 49 & 23.4 & Out \\
56524.91145\tnote{a} & -40694 $\pm$ 60 & 16.9 & In &     \hspace{10 mm}		& 56531.93561 & -40785 $\pm$ 43 & 25.7 & Out \\
56524.92455\tnote{a} & -40628 $\pm$ 70 & 17.2 & In &     \hspace{10 mm}		& 56531.94921 & -40802 $\pm$ 41 & 24.6 & Out \\
56524.93765\tnote{a} & -40757 $\pm$ 73 & 17.3 & In &     \hspace{10 mm}		& 56531.96477 & -40759 $\pm$ 46 & 24.9 & Out \\
56524.95077\tnote{a} & -40792 $\pm$ 66 & 18.6 & In &     \hspace{10 mm}		& 56531.98069 & -40720 $\pm$ 44 & 23.3 & In \\
56524.96386\tnote{a} & -40894 $\pm$ 65 & 17.5 & In &     \hspace{10 mm}		& 56531.99660 & -40733 $\pm$ 32 & 25.9 & In \\
56524.97697\tnote{a} & -40907 $\pm$ 71 & 17.6 & In &     \hspace{10 mm}		& 56532.01425 & -40752 $\pm$ 31 & 28.6 & In \\
56524.99006\tnote{a} & -40895 $\pm$ 70 & 17.6 & In &     \hspace{10 mm}		& 56532.03364 & -40806 $\pm$ 31 & 28.8 & In \\
56525.00316\tnote{a} & -40842 $\pm$ 63 & 17.6 & In &     \hspace{10 mm}		& 56532.05304 & -40821 $\pm$ 38 & 31.5 & In \\
56525.01626\tnote{a} & -40837 $\pm$ 61 & 18.5 & In &     \hspace{10 mm}		& 56532.07069 & -40864 $\pm$ 44 & 22.8 & In \\
56525.02936\tnote{a} & -40854 $\pm$ 66 & 18.4 & In &     \hspace{10 mm}		& 56532.08834 & -40875 $\pm$ 37 & 28.0 & In \\
56525.04246\tnote{a} & -40787 $\pm$ 62 & 17.6 & Out &     \hspace{10 mm}		& 56532.10773 & -40836 $\pm$ 38 & 26.9 & In \\
56525.05556\tnote{a} & -40619 $\pm$ 71 & 18.3 & Out &     \hspace{10 mm}		& 56532.12712 & -40797 $\pm$ 31 & 27.0 & In \\
56525.06866\tnote{a} & -40829 $\pm$ 62 & 17.9 & Out &     \hspace{10 mm}		& 56532.14651 & -40843 $\pm$ 44 & 28.4 & Out \\
56525.08176\tnote{a} & -40846 $\pm$ 45 & 20.1 & Out &     \hspace{10 mm}		& 56532.16590 & -40793 $\pm$ 39 & 30.9 & Out \\
56525.09485\tnote{a} & -40835 $\pm$ 72 & 18.8 & Out &     \hspace{10 mm}		& 56532.18529 & -40836 $\pm$ 35 & 29.9 & Out \\
56525.10796\tnote{a} & -40925 $\pm$ 63 & 18.6 & Out &     \hspace{10 mm}		& 56532.20295 & -40805 $\pm$ 46 & 26.4 & Out \\
\hline \\ [-2.5ex]
\end{tabular}
\label{table:RVs}
\begin{tablenotes}
\item [a] \textit{fast readout mode.}
\item [b] \textit{not used in analysis (high airmass).}
\end{tablenotes}
\end{threeparttable}%
\end{table*}

\begin{table*}
\centering
\begin{threeparttable}[b]
\caption{System parameters for HATS-3}
\centering
\begin{tabular}{l c}
\hline\hline \\ [-2.0ex]
Parameter & Value \\ [0.5ex]
\hline\hline \\ [-2.0ex]
{\textit{Parameters as given by \cite{2013AJ....146..113B}}} \\ [0.5ex]
{\textit{and used as priors in model}} \\
\hline \\ [-2.0ex]
Mid-transit epoch (2400000-HJD)\tnote{a}, $T_{0}$ & $56532.03799 \pm 0.00028$ \\
Orbital period\tnote{b}, $P$ & $3.547851 \pm 0.000005$ d \\
Semi-major axis\tnote{b}, $a$ & $0.0485^{+0.0004}_{-0.0006}$ AU \\
Orbital inclination\tnote{b}, $i$ & $86.2^{\circ} \pm 0.3^{\circ}$ \\
Impact parameter\tnote{b}, $b$ & $0.497^{+0.024}_{-0.027}$ \\
Transit depth\tnote{b}, $(R_{P}/R_{\star})^{2}$ & $0.01022 \pm 0.00060$ \\
Orbital eccentricity\tnote{c}, $e$ & 0.0 (assumed) \\
Argument of periastron\tnote{c}, $\varpi$ & N/A ($e=0$) \\
Stellar reflex velocity\tnote{c}, $K_{\star}$ & $125.7 \pm 15.7$ \mos\ \\
Stellar mass\tnote{c}, $M_{\star}$ & $1.209 \pm 0.036$ $M_{\odot}$ \\
Stellar radius\tnote{b}, $R_{\star}$ & $1.404 \pm 0.030$ $R_{\odot}$ \\
Planet mass\tnote{c}, $M_{P}$ & $1.071 \pm 0.136$ $M_{J}$ \\
Planet radius\tnote{b}, $R_{P}$ & $1.381 \pm 0.035$ $R_{J}$ \\
Stellar micro-turbulence\tnote{c}, $\xi_{t}$ & N/A \\
Stellar macro-turbulence\tnote{c}, $v_{mac}$ & N/A  \\
Stellar limb-darkening coefficient\tnote{c}, $q_{1}$ & 0.4135 (adopted) \\
Stellar limb-darkening coefficient\tnote{c}, $q_{2}$ & 0.3301 (adopted) \\
Velocity at published epoch $T_{P}$\tnote{c}, $V_{T_{P}}$ & $0.0$ \mos\ \\
RV offset between Bayliss and our complete dataset\tnote{b}, $V_{d}$ & $40792 \pm 16$ \mos\ \\
RV offset between 20 \& 27 Aug datasets\tnote{d}, $V_{s}$ & $144 \pm   20$ \mos\ \\ [0.5ex]
\hline\hline \\ [-2.0ex]
\textit{Parameters determined from model fit using} \\ [0.5ex]
\textit{our velocities from the complete data set} \\ [0.5ex]
\hline \\ [-2.0ex]
Projected obliquity angle, $\lambda_{T}$ & $3^{\circ} \pm 25^{\circ}$ \\ [0.5ex]
Projected stellar rotation velocity, ${v\sin i_{\star}}_{(T)}$ & $5.75 \pm 2.98$ \kms\ \\ [0.5ex]
\hline\hline \\ [-2.0ex]
\textit{Parameters determined from model fit using our velocities} \\ [0.5ex]
\textit{from the complete data set (errors empirically adjusted)} \\ [0.5ex]
\hline \\ [-2.0ex]
Projected obliquity angle, $\lambda_{T_{e}}$ & $4^{\circ} \pm 16^{\circ}$ \\ [0.5ex]
Projected stellar rotation velocity, ${v\sin i_{\star}}_{(T_{e})}$ & $5.75 \pm 1.70$ \kms\ \\ [0.5ex]
\hline\hline \\ [-2.0ex]
\textit{Parameters determined from model fit using Aug 20 velocities} \\ [0.5ex]
\hline \\ [-2.0ex]
Projected obliquity angle, $\lambda_{Aug20}$ & $-8^{\circ} \pm 55^{\circ}$ \\ [0.5ex]
Projected stellar rotation velocity, ${v\sin i_{\star}}_{(Aug20)}$ & $7.50 \pm 4.18$ \kms\ \\ [0.5ex]
\hline\hline \\ [-2.0ex]
\textit{Parameters determined from model fit using Aug 27 velocities} \\ [0.5ex]
\hline \\ [-2.0ex]
Projected obliquity angle, $\lambda_{Aug27}$ & $8^{\circ} \pm 33^{\circ}$ \\ [0.5ex]
Projected stellar rotation velocity, ${v\sin i_{\star}}_{(Aug27)}$ & $5.25 \pm 3.01$ \kms\ \\ [0.5ex]
\hline\hline \\ [-2.0ex]
\textit{Independent measurements of $v\sin i_{\star(Ind)}$ and} \\ [0.5ex]
{\textit{\cite{2013AJ....146..113B} $v\sin i_{\star(B)}$ published value}} \\ [0.5ex]
\hline \\ [-2.0ex]
Projected stellar rotation velocity, $v\sin i_{\star(Ind_{1})}$ & $5.2 \pm 0.6$ \kms\ \\ [0.5ex]
Projected stellar rotation velocity, $v\sin i_{\star(Ind_{2})}$ & $5.3 \pm 0.7$ \kms\ \\
[0.5ex]
Projected stellar rotation velocity, $v\sin i_{\star(B)}$ & $9.12 \pm {1.31}$ \kms\ \\ [0.5ex]
\hline 
\end{tabular}
\label{table:Parameters}
\begin{tablenotes}
\item [a] \textit{Parameter fixed from the transit photometry at the indicated value for final fit, but allowed to vary (as described in \S{3}) for uncertainty estimation.}
\item [b] \textit{Parameters fixed to the indicated value for final fit, but allowed to vary (as described in \S{3}) for uncertainty estimation.}
\item [c] \textit{Parameters fixed at values given by \cite{2013AJ....146..113B}.}
\item [d] \textit{Parameter fixed at value determined from the difference between $V_{d_{1}}$ \& $V_{d_{2}}$ from Aug 20 \& 27 datasets respectively.}
\end{tablenotes}
\end{threeparttable}%
\end{table*}

\section{Discussion}\label{sec:Discussion}
HATS-3 is the first exoplanetary system discovered from the HATSouth Transit Survey to have the Rossiter-McLaughlin effect measured. It is a relatively hot, $T_{eff}=6351 \pm 76$\,K, mid-F primary star \citep{2013AJ....146..113B} hosting a planet in a well-aligned orbit. This system joins the rapidly growing list of planetary systems for which spin-orbit alignments have been measured. A substantial fraction ($\sim41\%$) of the 76 systems with measured obliquities show spin-orbit misalignments ($\lambda > \frac{\pi}{8} = 22.5^{\circ}$)\footnote{\label{Rene_criteria}We have adopted Ren\'{e} Heller's criteria for misaligned orbits as given on the Holt-Rossiter-McLaughlin Encyclopedia; \url{http://www.physics.mcmaster.ca/~rheller/index.html}} and the majority of planets on high obliquity orbits are found around stars hotter than $T\geq6250$\,K (as noted by \citealt{2010ApJ...718L.145W}; \citealt{2012ApJ...757...18A}; and others). There are a few noteworthy exceptions to this general trend such as HAT-P-18b with $\lambda = 132^{\circ} \pm 15^{\circ}$ and $T_{eff}=4870 \pm 50$\,K \citep{2014A&A...564L..13E} and Kepler-63b with $\psi = {104^{+9}_{-14}}^{\circ}$ (true orbital obliquity $\psi$ as opposed to the sky-projected obliquity $\lambda$) and $T_{eff}=5576 \pm 50$\,K \citep{2013ApJ...775...54S}. 

Several mechanisms have been proposed to explain the high occurrence rate of exoplanetary systems observed to be in spin-orbit misalignment. These mechanisms include Kozai-Lidov cycles \citep{1962AJ.....67..591K,1962P&SS....9..719L,2007ApJ...669.1298F}, stellar internal gravity waves \citep{2013ApJ...772...21R}, chaotic star formation \citep[e.g.,][]{2011MNRAS.417.1817T}, primordial disk misalignments from interactions with a stellar binary \citep{2012Natur.491..418B,2014MNRAS.440.3532L}, planet-planet scatterings \citep{2008ApJ...686..580C}, and secular chaos \citep{2011ApJ...735..109W}. Misalignments produced through disk migration alone are disfavored because the disk from which planets form is expected to be well-aligned with the stellar spin axis of their host star. This assertion is well supported by recent observations of debris disks around nearby stars \citep[e.g.,][]{2013MNRAS.436..898K,2014MNRAS.438L..31G}. Since debris disks are material left behind from the formation of planetary systems \citep[e.g.,][]{2008ARA&A..46..339W}, the growing number of well aligned systems detected adds weight to the theory that most planetary systems form from protoplanetary disks that are aligned with their host star's equator. If a planet migrates solely through interaction with the disk, it is therefore expected to have its orbital plane remain aligned \citep{2010MNRAS.401.1505B}.

\begin{figure*}
\centering
\includegraphics[width=1.0\linewidth]{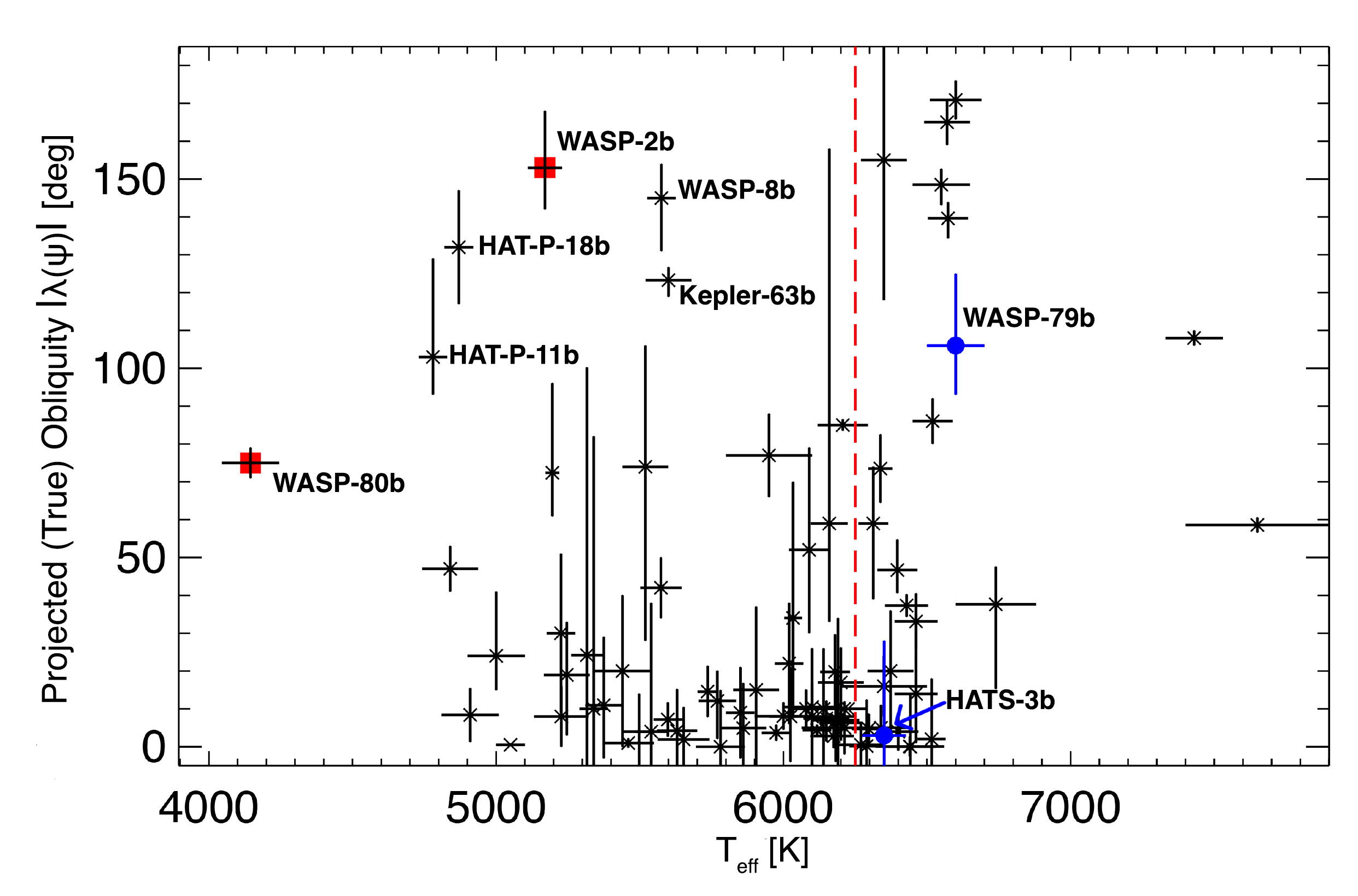}
\caption[LoF entry]{Projected orbital obliquity ($\lambda$) of exoplanets as a function of their host star's stellar effective temperature ($\mathrm{T_{eff}}$). This figure is updated from \citet{2014A&A...564L..13E}. The $T_{eff}=6250$\,K limit for which exoplanetary systems tend to display high obliquities is indicated by the red dashed line. HATS-3b and WASP-79b \citep{2013ApJ...774L...9A} have been included in this figure as the filled blue circles. The systems labeled to the left of the red-dashed line have anomalously large obliquities that break the observed trend of cool stars ($T_{eff}<6250$\,K) hosting planets on low obliquity orbits (WASP-80b, \citealt{2013A&A...551A..80T}; HAT-P-11b, \citealt{2010ApJ...723L.223W}; HAT-P-18b, \citealt{2014A&A...564L..13E}; WASP-2b, \citealt{2010A&A...524A..25T}; Kepler-63b, \citealt{2013ApJ...775...54S}; and WASP-8b, \citealt{2010A&A...517L...1Q}). The $\lambda$ values are not well constrained for the two planets (WASP-80b, \citealt{2013A&A...551A..80T}; and WASP-2b, \citealt{2011ApJ...738...50A}) marked in red squares.
\\ [+1.5ex]
(A color version of this figure will be available in the online journal.)}
\label{Figure:temp}
\end{figure*}

As the number of planetary systems with measured spin-orbit angles has grown, a few correlations have become apparent between the properties of the host star and the orbit of its planet. One of the first trends noted is that, as the temperature of the host star increases, so to does the likelihood of it hosting a planet on a significantly mis-aligned orbit. In particular, \citep{2010ApJ...718L.145W,2012ApJ...757...18A} observed that the measured obliquities fall into two distinct populations. Around the coolest stars ($T < 6250$\,K), the great majority of planets are found to be well aligned. In contrast, around the hotter stars ($T > 6250$\,K), the distribution of obliquities are far more random -- as can be clearly seen in Figure\,\ref{Figure:temp}. This dichotomy may be explained by the fact that stars $T_{eff}\geq6250$\,K have thin convective layers and are unable to realigned planets on high obliquity orbits through planet-star tidal interactions \citep{2012ApJ...757...18A}. These tidal interactions are thought to dampen orbital obliquites over time and primarily occur in the outer convective envelope of stars. Stars with effective temperature of $T_{eff}<6250$\,K have a much thicker convective envelope and can dampen obliquity more effectively on shorter timescales.

\begin{figure*}
\centering
\includegraphics[width=1.0\linewidth]{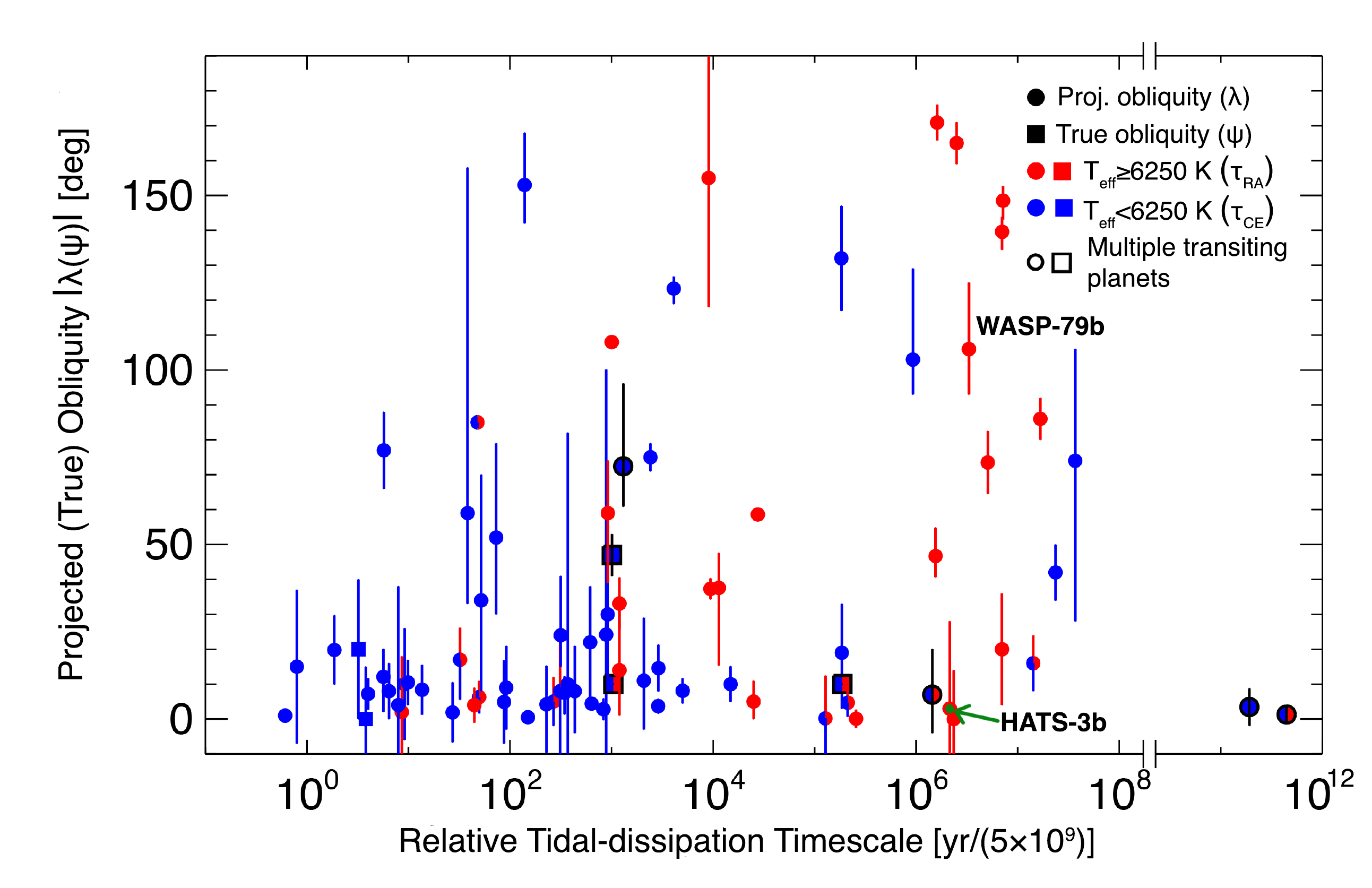}
\caption[LoF entry]{Projected orbital obliquity of exoplanetary systems as a function of their relative alignment timescale for stars with either convective (CE) or radiative envelopes (RA), calibrated from binary studies. This figure is updated from \citet{2013ApJ...771...11A}. The filled red circles with red error bars are for stars that have temperatures higher than $6250$\,K. The blue filled circles with blue error bars show stars with effective temperatures lower than $6250$\,K. The circles that are half red and blue show stars that have measured effective temperatures consistent with $6250$\,K from the $1\sigma$ interval. Multiple transiting planets are indicated by the dark black borders. Systems with measured projected obliquity ($\lambda$) are shown as circles while true obliquities ($\psi$) are shown as squares. We have included HATS-3b (indicated by the arrow and green dot) as well as WASP-79b from \citet{2013ApJ...774L...9A} in this figure.
\\ [+1.5ex]
(A color version of this figure will be available in the online journal.)}
\label{Figure:tidal}
\end{figure*}

In Figure\,\ref{Figure:tidal}, we show an updated plot of the projected orbital obliquity as a function of the relative tidal-dissipation timescales\footnote{Figure\,\ref{Figure:tidal} was produced from the compilation of stellar and planetary physical parameters as provided from \url{http://www.astro.keele.ac.uk/jkt/tepcat/allplanets-err.html}.} of exoplanetary systems as calibrated from binary studies and adopted from \citet{2013ApJ...771...11A} and \citet{2014arXiv1403.0652A}. We determined the tidal dissipation timescale, using the methods presented in \citet{2012ApJ...757...18A}, for HATS-3b as $\tau_{RA}=1.07\times10^{15}$\,yr (using the radiative timescale for alignment) or in relative terms $\tau_{RA}=2.13\times10^{6}$\,yr/($5\times10^{9}$\,yr). Taking into consideration the mass of the convective envelope (the second approach of \citealt{2012ApJ...757...18A}), we obtain $\tau_{mcz}=1.11\times10^{11}$\,yr and $\tau_{mcz}=35$\,yr/($3.2\times10^{9}$\,yr) if we normalize $\tau_{mcz}$ to the age of HATS-3. One can see from Figure\,\ref{Figure:tidal} that $\sim$2/3rds of exoplanetary systems have dissipation timescales shorter than that of HATS-3b. If HATS-3b did become misaligned during or shortly after migration, there would not have been enough time to realign the orbit. These results replicate those obtained in our earlier analysis of the mis-aligned system, WASP-79b \citep{2013ApJ...774L...9A}. For that system, we calculated the tidal dissipation timescales as $\tau_{RA}=3.3\times10^{15}$\,yr ($\tau_{RA}=6.6\times10^{6}$\,yr/($5\times10^{9}$\,yr)) and $\tau_{mcz}=1.60\times10^{11}$\,yr ($\tau_{mcz}=320$\,yr/($0.5\times10^{9}$\,yr)). Just as in the case for HAT-3b, these timescales are also sufficiently long that there has been insufficient time for a mis-aligned planet to re-align since the formation of the system.
 
In Figure\,\ref{Figure:temp}, we present an updated version of the projected orbital obliquity verses stellar temperature plot\footnote{Figure\,\ref{Figure:temp} was produced from the compilation of $\lambda$ and $\mathrm{T_{eff}}$ as provided from \url{http://www.astro.keele.ac.uk/jkt/tepcat/rossiter.html}.} as shown in \citet{2014A&A...564L..13E}, to which we have added our recently measured obliquities for WASP-79b and HATS-3b. At first glance, it appears that there is only a weak correlation between obliquity and effective temperature as there are several systems that are outliers. While this does indeed illustrate that cool stars can host planets in misaligned orbits, it also tells us that in order for them to host such planets, the tidal-dissipation timescale for realignment must be very long (as shown in Figure\,\ref{Figure:tidal}). Therefore a more relevant factor in determining whether a star can host a planet on a high obliquity orbit is the tidal-dissipation timescale and not just the stellar effective temperature. A long realignment timescale is possible for planets orbiting cool stars if the orbital distance-to-stellar radius ratio ($a/R_{\star}$) is sufficiently large and/or if the planet-to-star mass ratio ($M_{P}/M_{\star}$) is sufficiently small as the timescale is proportional to $a/R_{\star}$ and inversely proportional to $M_{P}/M_{\star}$. This supports the hypothesis, as proposed by \citet{2010ApJ...718L.145W} and \citet{2012ApJ...757...18A}, that whatever mechanism(s) are producing Hot Jupiters, are also randomly changing their spin-orbit angles and the systems with long realignment timescales will still have their initial (post-migration) spin-orbit angles while systems with short realignment timescales will have their spin-orbit angles realigned.

Giant planets are expected to form several AU away from their host star in the surrounding proto-planetary disk. This disk is expected to be well aligned with the star's spin-axis due to conservation of angular momentum from the collapse of the proto-stellar cloud \citep[e.g.,][]{1996Icar..124...62P,2011exha.book.....P,2013apf..book.....A}. During or shortly after formation, giant planets can migrate inwards (through various proposed migration mechanisms) to become Hot Jupiters where they have been observed to reside at separations as close as 0.01\,AU from their host star. In addition, nearly half of all Solar-type stars in the Milky Way have one or more stellar companions and planets have been found to be as likely to form around single stars as they are multiple stellar systems \citep[e.g.,][]{2010ApJS..190....1R,2012A&A...546A..10L,2012Sci...337.1511O,2013ApJ...768..127S,2014ApJ...783....4W}. This suggests that the stellar companions may have a significant role in shaping the evolution and migration of planets.

Several recent studies have suggested that Kozai resonances may explain the high frequency of spin-orbit misaligned exoplanets \citep[e.g., ][]{2007ApJ...669.1298F,2007ApJ...670..820W,2008ApJ...678..498N,2010A&A...517L...1Q,2011Natur.473..187N,2012PASJ...64L...7N,2013ApJ...769...86P}. Stellar companions to planet host stars that are widely separated (up to several hundred AU) and highly inclined with respect to the orbital plane of a planet can induce Kozai oscillations on the planet. These oscillations occur through the gravitational interactions between the planet and stellar companion and can drive the planet to high orbital obliquities. Eccentricity and orbital inclination are anti-correlated as described by the Kozai integral \citep[$I_{k}=\sqrt{1-e^{2}}\cos i$ from][]{1999ssd..book.....M}. The planetary orbital inclination is initially driven to match that of the orbital plane of the stellar companion. Then the Kozai oscillations will either cause the planetary eccentricity to increase while its orbital inclination (relative to the stellar companion) decreases or its orbital inclination (relative to the stellar companion) to increase while its eccentricity decreases. If the eccentricity is driven high enough, the planet will pass within a few stellar radii from its host star during periastron passage. The end result will be tidal dissipation and circularization of the orbit while the orbital inclination will be misaligned with the host star \citep{2008ApJ...678..498N}. This process may naturally explain the 3-day orbital period pile-up observed for Hot Jupiters produced through Kozai or secular chaos migration \cite{2011ApJ...735..109W}. \citet{2012arXiv1211.0554D} suggest, however, that high-eccentricity migration from Kozai resonances due to a stellar companion cannot be responsible for the production of all the Hot Jupiters. This is due to the lack of super-eccentric proto-hot Jupiters (planets currently undergoing high-eccentricity migration) discovered by Kepler. They suggest other migration processes, such as from disk migration, planet-planet scattering, planetary Kozai, or secular chaos, might instead be the dominant channel for the origin of Hot Jupiters. 

In addition, it is worth noting that such Kozai-driven eccentricity excitation will also increase the likelihood of the excited planet acting to destabilize the orbits of any other planets orbiting nearby. It is well established that, aside from certain resonant orbits \citep[e.g.,][]{2012ApJ...754...50R,2012ApJ...761..165W}, increases in the eccentricity of an exoplanet's orbit will act to decrease the stability of multi-planet systems \citep[e.g.,][]{2012ApJ...753..169W,2013ApJS..208....2W}. This might explain why no additional planets have been found orbiting the hosts of high-obliquity planets -- those planets were ejected as a part of the evolution of the high-obliquity planet to its current orbit. The one exception to this rule is Kepler-56, which hosts two planets in coplanar orbits that are misaligned with respect to the host's equator \citep{2013Sci...342..331H}. A massive companion in a wide orbit has been detected in this system and it is believed to be generating torques on the inner planets, driving them into coplanar orbits that are misaligned with the spin-axis of the host star \citep{2013Sci...342..331H}.

We have recently proposed to search for stellar companions around systems that host Hot Jupiters with measured obliquities to test the hypothesis that Kozai-Lidov cycles are the primary driver for spin-orbit misalignments \citep[as discussed in][]{2014arXiv1403.0652A}. Our search is being conducted by directly imaging a sample of nearby stars within 250\,pc using the Magellan Adaptive Optics (MagAO) and Clio2 infrared camera instruments on the 6.5\,m Magellan Telescope at the Las Campanas Observatory in Chile. We will be able to conclusively confirm or reject the presence of stellar companions to within 150\,AU in our sample and test if the Kozai mechanism is responsible for producing the majority of misaligned Hot Jupiters. A complementary survey to ours is being conducted by \citet{2014ApJ...785..126K} and they are searching for massive ($>1$\,$M_{J}$), long-period ($>1$\,yr) companions to close-in giant planets using radial velocity and adaptive optics imaging measurements. The first results from the \citet{2014ApJ...785..126K} program have found evidence for fifteen planetary and/or brown dwarf companions in fourteen systems (out of a total of 51 sampled systems) which suggests that the dynamical evolution of Hot Jupiters could be driven by distant, massive companions.

\section{CONCLUSION}
We have measured the spin-orbit alignment of the newly discovered Hot Jupiter HATS-3b, and find the planet's orbit to be well aligned to the projected rotational axis of its host star ($\lambda = 3^{\circ} \pm 25^{\circ}$). We obtained three separate values for the $v\sin i_{\star}$ of the stellar spin, namely: $v\sin i_{\star} = 5.75 \pm 2.98$\,\kms (from the RM effect measurements); $v\sin i_{\star} = 5.3 \pm 0.7$\,\kms (from a Gaussian fit to the least-squares deconvolution line profile); and $v\sin i_{\star} = 5.2 \pm 0.6$\,\kms\ (from a Gaussian fit to the cross-correlation function). They are all in good agreement with each other but in disagreement with the value of $v\sin i_{\star}=9.12 \pm 1.31$\,\kms\ from \citet{2013AJ....146..113B}, possibly due to extra broadening from macroturbulence. Nonetheless, we were able to robustly measure the spin-orbit angle of HATS-3b. Simultaneous photometry was obtained for a portion of the 27 August transit and was used to constrain the mid-transit time. Such photometry for future Rossiter-McLaughlin effect observations is vital to ensure proper monitoring of starspot-crossing events that can constrain the true obliquity of exoplanetary systems being studied.

One may expect HATS-3b to be in a misaligned orbit, given that its host star has $T_{eff}\geq6250$\,K\ and the realignment timescale for this system is very long ($\tau_{mcz}=1.11\times10^{11}$\,yr). Orbital obliquities likely are, however, initially distributed randomly from the migration processes that produce Hot Jupiters regardless of the value of $T_{eff}$ \citep{2010ApJ...718L.145W,2012ApJ...757...18A}. Therefore, we expect the observed obliquities to be randomly distributed for systems with long $\tau_{mcz}$ and low obliquites for systems with short $\tau_{mcz}$. This is true for almost all observed planetary systems for which spin-orbit angles have been measured, including HATS-3b, thus supporting the \citet{2012ApJ...757...18A} hypothesis. Alternatively, the low obliquity and short-period orbit of HATS-3b could just as well be explained by type I/II disk driven migration \citep[e.g.,][]{1996Natur.380..606L}. If this is indeed the case, then the orbit of HATS-3b was likely well-aligned to its host star's equator since its formation \citep[e.g.,][]{2010MNRAS.401.1505B}.

We are now beginning to unravel why high orbital obliquities are generally observed around stars with $T_{eff}\geq6250$\,K and long realignment timescales and not around stars with $T_{eff}<6250$\,K or short realignment timescales as more and more systems support the \citet{2012ApJ...757...18A} hypothesis. A lingering question remains, however, namely: \textit{what are the mechanism(s) most responsible for producing misaligned Hot Jupiters in the first place?} This important question will likely be resolved in the near future from an expansion of both the sample and the parameter space of spin-orbit alignment measurements. The parameter space least explored includes multiple planet and long-period transiting planet systems. Testing the various mechanism(s) thought to produce high obliquities, such as searching for stellar companions around stars hosting Hot Jupiters with obliquity measurements \citep[e.g.,][]{2014arXiv1403.0652A,2014ApJ...785..126K} and searching for evidence of additional planets orbiting the hosts of high-obliquity planets \citep[e.g.,][]{2014ApJ...785..126K}, will also be important avenues to pursue in resolving this mystery.

\acknowledgments

The research work presented in the paper at UNSW has been supported by ARC Australian Professorial Fellowship grant DP0774000, ARC LIEF grant LE0989347, ARC Super Science Fellowships FS100100046, and ARC Discovery grant DP130102695. Work at the Australian National University is supported by ARC Laureate Fellowship Grant FL0992131. We thank Jonathan Horner, Jonty Marshall, and Shaila Akhter for the helpful comments and suggestions on this manuscript. We acknowledge the use of the SIMBAD database, operated at CDS, Strasbourg, France. This research has made use of NASA's Astrophysics Data System, the Ren\'{e} Heller's Holt-Rossiter-McLaughlin Encyclopaedia (\url{http://www.physics.mcmaster.ca/~rheller/index.htm}), the Exoplanet Orbit Database and the Exoplanet Data Explorer at \url{exoplanets.org}, and the Extrasolar Planets Encyclopaedia at \url{http://exoplanet.eu}.

\bibliography{HATS-3b}

\end{document}